\documentstyle[epsfig,amsmath,aps,prl,twocolumn]{revtex}

\draft

\begin{document}
\twocolumn[\hsize\textwidth\columnwidth\hsize
\csname@twocolumnfalse%
\endcsname
\title{ The few-body problem for trapped bosons with large scattering length}
\author{S. Jonsell, H. Heiselberg, and C. J. Pethick}
\address{NORDITA, Blegdamsvej 17, DK-2100 Copenhagen \O, Denmark}
\date{\today}

\maketitle

\begin{abstract}
We calculate energy levels of two and three bosons trapped
in a harmonic oscillator potential with oscillator length
$a_{\mathrm  osc}$. The atoms are assumed to interact through a
short-range potential with a scattering length $a$, and the short-distance
behavior of the three-body wave function is characterized by a parameter
$\theta$. For large positive $a/a_{\mathrm osc}$, the energies of
states which, in the absence of the trap, correspond
to three free atoms approach values independent of $a$ and $\theta$.
For other states
the $\theta$ dependence of the energy is strong, but
the energy is independent of $a$ for $|a/a_{\mathrm osc}|\gg1$.
\end{abstract}

]

By exploiting the properties of Feshbach resonances, it has become
possible to tune atomic scattering lengths to essentially any value,
either repulsive or attractive \cite{feshbach}. This has led to
renewed interest in the properties of systems with large scattering
length $a$, a problem which has also been studied extensively in the context
of nuclear physics, because the scattering lengths for nucleon-nucleon
interactions are large in magnitude compared with the spatial scale
$r_0$ of the interparticle interaction.

A novel possibility for exploring experimentally the properties of
few-body systems is to trap atoms in a periodic potential created by
standing-wave light beams, a so-called optical lattice.  For example,
in the experiments described in Ref.\ \cite{bloch}, up to
$200,000$ $^{87}$Rb atoms were distributed over more than $150,000$ lattice
sites, thereby creating a large number of few-body systems with up to 2.5
atoms each on average.
This provides one motivation for theoretical studies of few-body
problems in traps.  A second motivation is  to understand theoretically the
properties of bulk matter when the magnitude of the scattering length is
large compared with the interparticle spacing $~n^{-1/3}$, where $n$ is the
particle density.  In this regime, the usual low density
approximations, which are valid only if $n|a|^3 \ll 1$, fail. Recently,
one has sought a universal description, with $n|a|^3$ as the only
parameter, of the many-body problem for  Fermi
\cite{heis01} and Bose \cite{cowe01} systems interacting via short-range
potentials ($r_0\ll n^{-1/3}$), but with $n|a|^3\simeq 1$. Studies of
three-body recombination in ultracold Bose gases have, however, revealed that
the rate depends on an additional three-body parameter \cite{recombination},
and in a recent work, it has been suggested that the equation of state of a
Bose gas in the limit $n^{-1/3}\ll |a|$ may depend on the three-body
parameter \cite{braaten}. Study of the few-body system in a trap is therefore
of interest in order to obtain insights into the properties of bulk matter.

In this Letter we calculate the energy of systems with two or three
identical bosons of mass $m$ in a harmonic trapping potential
$V_{\mathrm trap}=m\omega^2r^2/2$, where $\omega$ is the frequency
of oscillations of a particle about the minimum of the potential.  The length
associated with the zero-point motion of a particle in the trap is
$a_{\mathrm osc} =\sqrt{\hbar/m\omega}$. The bosons are assumed to interact
via a short-range potential.  To take this into account, we impose the boundary condition
$\partial \ln(r_{ij}\Psi)/\partial r_{ij}|_{r_{ij}\rightarrow0}=-1/a$. Here $\Psi({\bf r}_1, \ldots {\bf r}_N)$
is the wave function of the $N$-body system, ${\bf r}_i$ being the
coordinates of the particles, and ${\bf r}_{ij}={\bf
r}_i - {\bf
r}_j$.

We first consider the two-body problem.
Solving the Schr\"odinger equation for two bosons in a
harmonic oscillator potential, one finds that the eigenenergies $E$,
excluding the center
of mass contribution, are given by \cite{busc98}
\begin{equation}
\frac{a}{a_{\mathrm
osc}}=\frac{1}{\sqrt{2}}\frac{\Gamma[1/4-E/(2\hbar\omega)]}
{\Gamma[3/4-E/(2\hbar\omega)]}.
\label{eq:E2b}
\end{equation}
 For $a=0$
the $s$-wave states have energies $E=\hbar\omega(2\nu+3/2)$, ($\nu=0,1,\dots$)
at which the gamma function in the
denominator diverges, while for
$a/a_{\mathrm osc} \rightarrow\pm\infty$, the energies
are $E=\hbar\omega(2\nu+1/2)$.
Hence, when the bosons are confined to a volume smaller than $|a|^3$, the
scattering length becomes unimportant.  Interestingly, the energy does not
increase without limit as $a$ becomes large, as one would predict from the
standard low-density result for the energy.

We now investigate the analogous problem for three trapped bosons, and
adopt the
method of Ref.\ \cite{fedo93}. The wave function of three identical
bosons may be decomposed in the usual Faddeev fashion as
\begin{equation}
  \label{eq:faddeev}
  \Psi({\bf r}_1,{\bf r}_2,{\bf r}_3)=\psi({\bf x},{\bf y})+
  \psi({\bf x}',{\bf y}')+\psi({\bf x}'',{\bf y}'').
\end{equation}
Here the Jacobi coordinates ${\bf x}={\bf r}_{12}$, and ${\bf y}
={\bf r}_3-({\bf r}_1+{\bf r}_2)/2$ have
been used. If interactions between atoms are purely
$s$-wave, $\psi$ is a function of the
hyperradius $\rho=(x^2/2+2y^2/3)^{1/2}$ and the hyperangle
$\alpha=\arctan(\sqrt{3}x/2y)$ only. Variables related to the original
ones by cyclic permutations of the particles are denoted by primes and
double primes.

The wave function $\psi$ may be represented exactly by an infinite sum
\begin{equation}
\label{eq:expansion}
\psi(\rho,\alpha)=\frac{1}{\sqrt{3}}\sum_{i}\frac{f_{i}(\rho)}{\rho^{5/2}}
\frac{\phi_{i}(\rho,\alpha)}{\sin\alpha\cos\alpha},
\end{equation}
where $\phi_{i}(\rho,\alpha)$ are eigensolutions to the
hyperangular part of the problem, which for a short-range potential
reduce to
\begin{figure}[t]
  \begin{center}
    \epsfig{figure=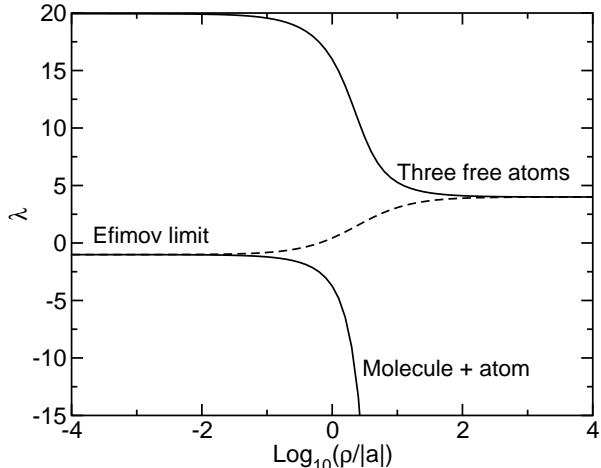,width=8.5cm}
    \caption{Eigenvalue $\lambda_i(\rho/|a|)$ of the hyperangular equation
      for $a>0$ (solid) and $a<0$ (dashed). }
    \label{fig:lamall}
  \end{center}
\end{figure}
\vspace{-0.6cm}
\begin{equation}
  \label{eq:phi}
  \phi_{i}(\rho,\alpha)=\sqrt{N(\rho)}\sin\left[
    \sqrt{\lambda_i}(\alpha-\pi/2)\right].
\end{equation}
The eigenvalues $\lambda_i$ are determined by the boundary
condition as $\alpha\rightarrow0$, i.e. when two of the bosons
approach each other, and are given by the equation
\begin{equation}
  \label{eq:lambda}
  \sqrt{\lambda_i}\cos\left[\sqrt{\lambda_i}\frac{\pi}{2}\right]-
  \frac{8}{\sqrt{3}}\sin\left[\sqrt{\lambda_i}\frac{\pi}{6}\right]=
  \sqrt{2}\frac{\rho}{a}\sin\left[\sqrt{\lambda_i}\frac{\pi}{2}\right].
\end{equation}

The hyperradial wave functions $f_{i}$ are the solutions of
an infinite number of coupled differential equations
\begin{multline}
\label{eq:system}
  \left(-\frac{\partial^2}{\partial\rho^2}+
  \frac{\lambda_i(\rho/a)-1/4}{\rho^2}-Q_{ii}(\rho)+
  \frac{\rho^2}{a_{\mathrm osc}^4}-\frac{2mE}{\hbar^2}\right)f_{i}(\rho)\\
  =\sum_{j\neq i}\left(Q_{ij}(\rho)+2P_{ij}(\rho)
\frac{\partial}{\partial\rho}\right)f_{j}(\rho).
\end{multline}
The effective potential for the
hyperangular motion therefore depends on the eigenvalue $\lambda_i$ associated
with the hyperangular degrees of freedom. The coupling matrices are given by
the kinetic-energy operator acting on the hyperangular wave functions:
\begin{equation}
\label{eq:PQ}  
P_{ij}(\rho)=\left\langle\Phi_i\left|  
\frac{\partial}{\partial\rho}\right|\Phi_j\right\rangle,\,  
Q_{ij}(\rho)=\left\langle\Phi_i\left|  
\frac{\partial^2}{\partial\rho^2}\right|\Phi_j\right\rangle,  
\end{equation}
where $\Phi_{i}(\rho,\alpha,\alpha',\alpha'')=
\phi_{i}(\rho,\alpha)+\phi_{i}(\rho,\alpha')+ 
\phi_{i}(\rho,\alpha'')$. An approximation which has worked well in a
number of applications \cite{niel01} is to neglect the coupling terms $P$
and $Q$. This is an adiabatic approximation in
the sense that the hyperangular functions
$\Phi_{i}(\rho,\alpha,\alpha',\alpha'')$ are assumed to vary much more
slowly with hyperradius than with the hyperangles. The method can be
refined by including coupling between a finite subset of states.

In Fig.\  \ref{fig:lamall} we plot $\lambda_i(\rho/|a|)$
calculated from Eq.\ (\ref{eq:lambda}) for some 
adiabatic states. For both $a>0$ and $a<0$ we show the state
approaching $\lambda_i=4$ for large $\rho/|a|$.
This would, in the absence of
the trap, correspond at large $\rho$ to three free atoms, and it is the
state of greatest interest in the present work.  Also shown for $a>0$ is the
result for the state which would correspond to a diatomic molecule and an
atom.

Let us now examine the form of the adiabatic hyperspherical wave
functions. The functions of interest are ones which remain finite as
$\rho \rightarrow 0$. The effective potential for small $\rho/|a|$
varies as $(\lambda_i(0) -1/4)\rho^{-2}$. Consequently for
$\lambda_i(0) >0$ the hyperradial wave function $f_i$ tends to zero
as $\rho^{\sqrt{\lambda_i(0)}+1/2}$  for small
$\rho/|a|$, and it does not depend on the nature of three-body
correlations when
no two atoms are separated by more than the typical range of
the interaction. The situation is quite different for $\lambda_i(0) <0$,
since the effective potential is then attractive at small $\rho$. The
two linearly independent solutions oscillate rapidly as $\rho\rightarrow 0$,
but do not diverge, and the general solution for small $\rho$ may be written
\begin{equation}
  f(\rho)\rightarrow \sqrt{\rho}
                \sin[|s_{0}|\ln(\rho/a_{\mathrm osc})+\theta],
        \label{eq:frho0}
\end{equation}
where $s_{0}=\sqrt{\lambda_0(0)}\simeq1.00624i$ and $\theta$ is a three-body parameter
analogous to the phase shift for the two-body problem.

\begin{figure}[t]
  \begin{center}
    \epsfig{figure=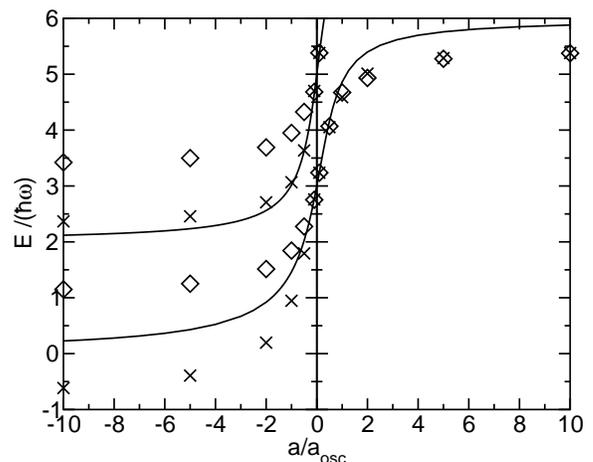,width=8.5cm}
    \caption{Three-body energies as a function of $a/a_{\mathrm osc}$,
      for $\theta({\rm mod}\,\pi)=0$ (crosses) and $\theta({\rm mod}\,\pi)=\pi/2$ (diamonds). For comparison the solid line shows the energy obtained when three-body
correlations are neglected.}
    \label{fig:Eofa}
  \end{center}
\end{figure}

We turn now to the energy eigenvalues.  In Fig.\ \ref{fig:Eofa} we exhibit
numerical results for the solution of Eq.\ (\ref{eq:system}), where the only
states included are those shown in Fig.\ \ref{fig:lamall}.
For $a>0$, coupling between the two hyperangular functions has been
included. Since this is weak, it is still possible to separate
states that for large $\rho$ correspond to three free atoms from states
that correspond to a molecule and an atom.
We show the energy only of the states that for large $\rho$
correspond to three free atoms, since these are the states of interest here.

\vspace{0.5cm}

For small $|a/a_{\mathrm osc}|$ the energy of the lowest harmonic
oscillator state can be expanded as
\begin{equation}
E=\hbar\omega\left[3+\frac{6}{\sqrt{2\pi}}\frac{a}{a_{\mathrm osc}}+
c\left(\frac{a}{a_{\mathrm osc}}\right)^2+\dots\right].
        \label{eq:Elow}
\end{equation}
A lower bound to the energy may be found by omitting
$Q_{ii}(\rho)$ in Eq.\ (\ref{eq:system}).  An upper bound may be
obtained by evaluating the energy for the trial wave function $
f_i^{(0)} \Phi_i$, where  $ f_i^{(0)}$ is the solution of
Eq.\ (\ref{eq:system}) in the absence of the coupling terms \cite{star79}.
Including the same states as before, we find $-0.55<c<-0.25$. For comparison,
if three-body correlations are neglected, the correlation energy (the
difference between the total energy and that of the corresponding state in the
absence of interactions) is three times the correlation energy for two bodies,
as obtained from Eq.\ (\ref{eq:E2b}).  The resulting approximation for the
energy has the same term linear in $a/a_{\mathrm osc}$, but the quadratic
term has a different coefficient, $c=6(1-\ln2)/\pi\simeq0.59$. Hence, three-body
effects are to leading order proportional to  $(a/a_{\mathrm osc})^2$.

With increasing $|a|/a_{\rm osc}$ the dependence of the energy on $\theta$
grows.  For $a<0$ this is due to the sensitivity of the adiabatic state to
the inner boundary condition.  The adiabatic state for $a>0$ is insensitive
to $\theta$, and the $\theta$ dependence arises due to coupling to states
having different hyperangular functions. Energies for $\theta=0$ and
$\theta=\pi/2$ are compared in Fig.\ \ref{fig:Eofa}, and the difference
grows with $a/a_{\mathrm osc}$ until $a/a_{\mathrm   osc}\sim1$, where it is
$\sim5\%$ of the correlation energy. At larger values of $a/a_{\mathrm osc}$
the $\theta$ dependence decreases. This is an effect of the confining
potential. The hyperradial wave function $f(\rho)$ varies on a scale $\sim
a_{\mathrm osc}$. As shown above, the hyperangular wave function
$\Phi_{i}$ does, however, depend on the hyperradius through the
combination $\rho/a$. Hence the couplings $P$
and $Q$ Eq.\ (\ref{eq:PQ}) are suppressed by factors $a_{\mathrm
  osc}/a$ and $(a_{\mathrm osc}/a)^2$ respectively when $a>a_{\mathrm
  osc}$.

\begin{figure}[tbh]
  \begin{center}
    \epsfig{file=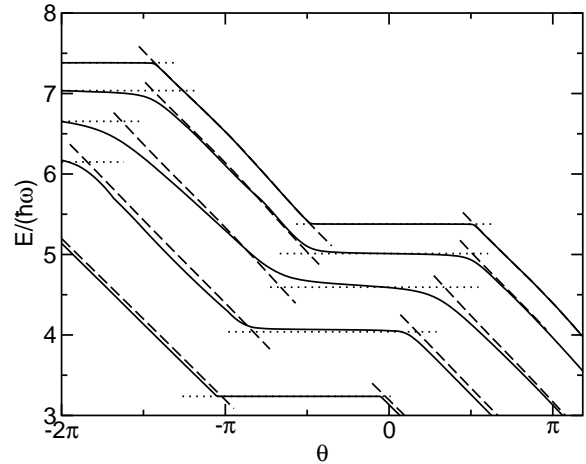,width=8.5cm}
    \caption{Energy of $\theta$ for (from below) $a/a_{\mathrm osc}=$ 0.1,
      0.5, 1, 2, and 10. Dashed and dotted lines show
      energies in the adiabatic approximation, i.e.\ neglecting
      couplings between hyperangular functions. $\theta$ has been extended over
      several branches $\pi$ in order to simultaneously show several
      physical states (see text).}
    \label{fig:theta}
  \end{center}
\end{figure}

The $\theta$ dependence of the energy is shown in more detail in Fig.\
\ref{fig:theta} for $a>0$. Physical conditions only determine
\mbox{$\theta({\rm mod}\,\pi)$},
but to exhibit the structure we have continued the energy curves over a
larger range of $\theta$.
Hence, given values of $a/a_{\mathrm osc}$  and $\theta$, $E(\theta)$,
$E(\theta+\pi)$, $E(\theta+2\pi)$ etc.\ along the same curve
represent the energies of different physical states. States corresponding
mainly to free atoms at large $\rho$ are represented by the nearly
$\theta$-independent segments of the curves. In the adiabatic approximation,
i.e.\ neglecting all couplings, these states have energies shown by the
dotted lines. Energies of other states, i.e.\ states associated with
molecular bonding, are given by the dashed lines.
The difference between the solid and dashed/dotted
lines is the effect of the couplings between the two hyperangular functions.
The couplings are most important when $a\simeq
a_{\mathrm osc}$, and are almost entirely absent for $0<a<0.1a_{\mathrm
  osc}$ and $a>10a_{\mathrm osc}$. The results for $a=10 a_{\mathrm
  osc}$ indicate that for $a\gg a_{\mathrm osc}$ the couplings vanish
and our solution becomes exact.

\vspace{-0.5cm}

In the limit $a/a_{\mathrm osc} \rightarrow\pm\infty$ Eq. (\ref{eq:system})
may be solved
exactly. The energy
tends to a constant value, which for the lowest state with a given
$\lambda_i(0) >0$
 is $E=\hbar\omega(1+\sqrt{\lambda_i(0)})$. For the state which at large $\rho$
corresponds to free atoms, and for $a>0$, one has $\lambda_i(0) \simeq 19.94$
and $E\simeq5.47\hbar\omega$. The
corresponding energy if only two-body correlations are included is
$E=6\hbar\omega$, and therefore the correlation energy is reduced by
approximately $18\%$ by three-body effects. For $\lambda_i(0) <0$ the energies are
functions of the boundary  condition for
$\rho\rightarrow0$, i.e.\ of $\theta$, and are given by the solutions of
the equation
\begin{equation}
  \theta=-\arg\left\{\frac{\Gamma[1/2-E/(2\hbar\omega)-s_{0}/2]}{\Gamma[1-s_{0}]}\right\}.
        \label{eq:Eainfinite}
\end{equation}
For $E>0$ this equation gives harmonic oscillator states shifted by
the interaction, with spacings $E_{n+1}-E_{n}\simeq2\hbar\omega$,
while for $E<0$ it gives bound three-body states, so-called Efimov states
with energies scaling like $E_{n+1}/E_n=\exp(2\pi/|s_0|)\simeq515$
\cite{efim70} perturbed by the oscillator potential. For
example $\theta=0$ gives energies $E
\simeq \dots$, $-291649\hbar\omega$,
$-566\hbar\omega$, $-0.85\hbar\omega$, $2.27\hbar\omega$,
$4.48\hbar\omega$, $\dots$ .

Let us now examine the implications of our results for the properties of bulk
matter. On physical grounds, we would expect the energy per particle in
trapped few-body systems to be similar to that for bulk matter if the
oscillator length is replaced by the typical particle separation.   
For $|a| \gg a_{\rm osc}$, the energy per particle of two- and
three-body states  for which $\lambda_i(0) >0$ is $\sim \hbar \omega$ greater
than in the absence of interactions.  Since the density of particles is $n\sim
1/a_{\rm osc}^3$, this corresponds to  an energy per particle  $\sim(\hbar^2/m)
n^{2/3}$.  There is no dependence on $\theta$ for these states. The state
corresponding to unbound atoms in a trap falls into this class for
positive $a$. This suggests that the energy per
particle of bulk matter has the universal dependence $\sim Cn^{2/3}$, where
$C$ is a constant that does not depend on the three-body parameter
$\theta$.  In particular, it does not have density-dependent contributions
that are periodic in $\ln n$, contrary to the possibility suggested
in Ref.\cite{braaten}. There are other states, such as those
associated with molecular states bound by the interatomic potential, which do
exhibit a strong dependence on $\theta$. These latter states are the final
states in recombination processes, where diatomic molecules are formed from
free atoms, and consequently recombination rates depend strongly on $\theta$.

The expansion Eq.\ (\ref{eq:Elow}) of the energy of three trapped
bosons, may be compared to the low-density expansion for the energy of
$N$ bosons in a homogenous gas
\cite{leeyang} \begin{equation}
  \label{eq:EN}
  E=\frac{2\pi \hbar^2 a n}{m}(N-1)\left[
  1+\frac{128}{15}\left(\frac{na^3}{\pi}\right)^{1/2}+\dots\right].
\end{equation}
For comparison, our result  Eq.\ (\ref{eq:Elow}) for the energy of
three trapped bosons with $|a/a_{\mathrm  osc}|\ll1$, may be rewritten as
\begin{equation}
E=3\hbar\omega+\frac{2\pi \hbar^2  a \bar{n}}{m}(N-1)\left[1+
\frac{c\sqrt{2\pi}}{6}\frac{a}{a_{\mathrm osc}}+\dots\right].
        \label{eq:Elow2}
\end{equation}
Here the average density, weighted by the particle number, is defined by
$\bar{n}=  \int n({\bf r})dN({\bf r})/ \int dN({\bf r}) =  \int n^2({\bf
r})d{\bf r}/N$, where $ dN({\bf r})=n({\bf r})d{\bf r}$, and therefore for a
homogeneous condensate $\bar{n}=n$. The three-body and many-body results agree
for the term linear in $a$, which usually is the only term retained
in,  e.g.\ the 
Gross-Pitaevskii equation for condensates. Beyond the linear term the two
expansions differ. The next term in Eq.\ (\ref{eq:EN}) is of order
$(na^3)^{1/2}$ relative to the first one, 
while in Eq.\ (\ref{eq:Elow2}) it is proportional to $a/a_{\mathrm
osc}$. If $N\lesssim a_{\mathrm osc}/a$ the most important correction to the
linear dependence of energy on $a$ will be the second term in Eq.\
(\ref{eq:Elow2}), which arises from the trapping potential.

To determine when the three-body parameter $\theta$ becomes important, we
note that, in the absence of the trap, the S-matrix element between the
adiabatic state corresponding to free atoms and that corresponding to a
diatomic molecule and an atom is proportional to $(ka)^2$,
where $k$ is the wave number of the colliding atoms.
Therefore the recombination rate is proportional to $a^4$,
in agreement with earlier work  \cite{recombination}.
The constant of  proportionality depends on $\theta$, because of the $\theta$
dependence of the final state.
The same result is valid 
in a trap with $a_{\mathrm osc}\gg|a|$, where due to the
zero point motion $k$ is of order $1/a_{\mathrm osc}$.
Therefore, the first term
 in the small $a/a_{\mathrm osc}$ expansion of the energy 
[Eq.\ (\ref{eq:Elow})] that depends explicitly on the
coupling between different adiabatic states (and therefore on
$\theta$) is of order  $(a/a_{\mathrm osc})^4$. 
In the expansion Eq.\ (\ref{eq:Elow2}) this
corresponds to the cubic term, since there one power of $a/a_{\mathrm osc}$
has been factored out.
To this order the result agrees with the results for the energy density of
the homogeneous Bose gas in Ref.\ \cite{braa99}.  
As we saw above, for $a \gg a_{\mathrm osc}$ the couplings $P$ and $Q$ between
hyperangular functions are {\em reduced}, which will tend to suppress
recombination rates.   

In conclusion,
we have found that there are states of the trapped three-boson system with
very weak dependence of the energies on the
three-body parameter $\theta$. This dependence vanishes completely
in the limit of an infinitely large scattering length. 
The fact that the energy of these states is independent of the three-body
parameter in the limit of large scattering lengths gives some grounds to hope
that the energy of trapped systems with a larger number of bodies is
insensitive to the parameters describing the four- and higher-body problems,
and that similar conclusions apply to the energy of bulk matter.  
Topics for future study include higher-body correlations, and 
the lifetime of a condensate with a large scattering length. 

We are grateful to Eric Braaten and Oleg I. Kartavtsev for valuable discussions.

\end{document}